# Spin-Orbit Torque Engineering in β-W/CoFeB Heterostructures via Ta and V Alloying at Interfaces

Gyu Won Kim[1†], Do Duc Cuong[2†], Yong Jin Kim[1], In Ho Cha[1], Taehyun Kim[1], Min Hyeok Lee[1], OukJae Lee[3], Hionsuck Baik[4], Soon Cheol Hong[2], Sonny H. Rhim[2*], and Young Keun Kim[1*]

Correspondence: Young Keun Kim (ykim97@korea.ac.kr) or Sonny H. Rhim (sonny@ulsan.ac.kr)

[1]Department of Materials Science and Engineering, Korea University, Seoul 02841, Korea.

[2]Department of Physics, University of Ulsan, Ulsan 44610, Korea.

[3]Center for Spintronics, Korea Institute of Science and Technology, Seoul 02792, Korea.

[4]Seoul Center, Korea Basic Science Institute (KBSI), Seoul, 02841, Korea

A full list of author information is available at the end of the article

[†] These authors contributed equally: Gyu Won Kim, Do Duc Cuong

**Abstract**

Spin-orbit torque manifested as an accumulated spin-polarized moment at nonmagnetic normal metal and ferromagnet interfaces is a promising magnetization switching mechanism for spintronic devices. To fully exploit this in practice, materials with a high spin Hall angle, i.e., a charge-to-spin conversion efficiency, are indispensable. To date, very few approaches have been made to devise new nonmagnetic metal alloys. Moreover, new materials need to be compatible with semiconductor processing. Here we introduce W-Ta and W-V alloys and deploy them at the interface between β-W/CoFeB layers. First, spin Hall conductivities of W-Ta and W-V structures with various compositions are carried out by first-principles band calculations, which predict the spin Hall conductivity of the W-V alloy is improved from $-0.82 \times 10^3$ S/cm that of W to $-1.98 \times 10^3$ S/cm. Subsequently, heterostructure fabrication and spin-orbit torque properties are characterized experimentally. By alloying β-W with V at a concentration of 20 at%, we observe a large enhancement of the absolute value of spin Hall conductivity of up to $-(2.77 \pm 0.31) \times 10^3$ S/cm. By employing the X-ray diffraction and scanning transmission electron microscopy, we further explain the enhancement of spin-orbit torque efficiency is stemmed from W-V alloy between W and CoFeB.

**Introduction**

Recent decades have witnessed tremendous progress in spintronics, where spin-orbit torque (SOT)[1,2], which is much faster and more energy-efficient than spin-transfer-torque (STT)[3,4], has attracted attention technologically and scientifically. SOT is the key ingredient of magnetization switching, enabling devices with magnetic memories[5,6] and logic[7,8]. In typical nonmagnetic metal (NM)/ferromagnetic metal (FM) structures, the polarized spin moments, which are carried by the spin current and accumulated at the NM and FM interface, manifest the SOT. Accordingly, spin-Hall angle ($\theta_{SH}$), a ratio of the amount of generated spin current relative to injected charge current, is widely described as the SOT efficiency.

It is known that the generation of spin current is due to spin-orbit coupling in the magnetic heterostructure[9,10]. Although it has recently been found that spin current can be generated in the FM (e.g., NiFe)[11], still, the major contribution of SOT results from the spin current generated in the NM. Thus, several efforts to increase $\theta_{SH}$ have been explored in various classes of materials, including topological insulators and Weyl semimetals[12-14]. These materials exhibit relatively large $\theta_{SH}$ values. However, it is not easy to apply them to actual semiconductor manufacturing processes due to the complexity of the growth method and lack of thermal stability. Therefore, it is important to design NM with manufacturing process-friendly materials, such as W, widely used in the current semiconductor-industry. It is also equally essential to find material combinations that possess perpendicular magnetic anisotropy (PMA), which is the key to achieving high bit density[15,16].

As semiconductor-process friendly materials, high resistive β-W are good candidates, which shows PMA with CoFeB and large SOT efficiency $\theta_{SH}$ = -0.33 ~ -0.40[17,18].

More efforts have been made to enhance $\theta_{SH}$ in β-W derivatives. One of them modifies the sputtering atmosphere to fabricate the $WO_x$[19] layer, and the other adapts multistep depositions to make the thick β-W up to 16 nm[20], both of which reach $\theta_{SH}$ = -0.5. In the former study, the PMA strength became weaker with oxygen incorporation, while in the latter case, the processing time increased. Moreover, Ta substitution into β-W has been theoretically predicted to give $\theta_{SH}$ = -0.5, where the intrinsic band structure plays such a role[21].

Theoretical work is extended here: since V has the same number of valence electrons as Ta, presumably V is expected to play a similar role in A15 structured β-W. Thus, Ta and V substitution in β-W have been explored for a whole range of alloy concentrations. All possible alloy configurations are considered for each concentration by substituting the A15-ordered W atomic site to Ta or V.

Based on our first-principles calculations, we introduced an interfacial layer, $W_{100-x}X_x$ (where X is either Ta or V), in between β-W and CoFeB layers. We fabricated β-W/$W_{100-x}X_x$/CoFeB/MgO/Ta heterostructure film stack on Si wafers for various alloy compositions, 0 < x < 100 at%. Due to the presence of bottom β-W layers, we expected to maintain the β-phase in the $W_{100-x}X_x$ interfacial layers as much as possible. Most of the composition range showed the PMA, and the harmonic Hall measurement confirmed that the $\theta_{SH}$ increased up to -0.49, for an approximate 40% enhancement over β-W, which agreed well with our calculated results. Such large SOT was further crosschecked by other measurement methods such as domain wall depinning and the propagation model[22].

**Materials and methods**

**Theoretical calculations**

First-principles calculations were performed using the Vienna Ab Initio Simulation Package (VASP)[23] with projector augmented-wave (PAW) basis[24]. An energy cutoff of 500 eV was chosen along with a 16 × 16 ×16 *k* mesh for summation in the Brillouin zone. Generalized-gradient approximation (GGA) was employed for the exchange-correlation potential as parametrized by Perdew, Burke, and Ernzerhof (PBE)[25]. The spin Hall conductivity ($\sigma_{SH}$) was calculated using the Kubo formula in the linear response theory[26]:

$$\sigma_{SH} = \frac{e}{\hbar} \sum_{kn} f_{kn}\, \Omega_n^z(\boldsymbol{k}), \tag{2}$$

where $f_{kn}$ is the Fermi-Dirac function for the n-th band at $\boldsymbol{k}$, so that the Berry curvature of the n-th band at $\boldsymbol{k}$ is expressed as:

$$\Omega_n^z(\boldsymbol{k}) = -2\, Im \sum_{n' \neq n} \frac{\langle kn|j_x|kn'\rangle \langle kn'|v_y|kn\rangle}{\left(e_{\boldsymbol{k},n} - e_{\boldsymbol{k},n'}\right)^2}, \tag{3}$$

where $j_x = \frac{\beta}{2}\{\Sigma, v_x\}$ is the spin current and $\Sigma$ is the spin operator in full relativistic formalism. Here, the interpolation technique[27,28] using Wannier90 was employed with *s*, *p*, and *d* orbitals of V, W, and Ta atoms. Throughout this work, we considered the xy component of the spin Hall conductivity with the spin axis along the *z*-direction. For alloys, the spin Hall conductivity of concentration *x* was calculated for possible configurations, which is thermodynamically averaged using the Boltzmann factor by taking the relative energies into account (Supplementary Note 1).

**Sample preparation**

Samples were sputtered onto 1.25 × 1.25 cm$^2$ thermally oxidized Si wafers under a base pressure below $5 \times 10^{-9}$ Torr. The thickness of the Si-oxide layer was 300 nm. The stacking structure of the samples was W (4)/W$_{100-x}$X$_x$ (2)/CoFeB (0.9)/MgO (1)/Ta (2) (numbers represent the thickness in nm), where X is either Ta or V. We employed a 4 nm thick β-W layer to maintain the β-phase in the 2 nm thick W$_{100-x}$X$_x$ alloy layers. The alloy interfacial layer composition was varied in every 10 at% steps by changing the sputtering power densities of W and Ta (or V) targets during co-deposition. Here, the composition of the CoFeB target was $Co_{40}Fe_{40}B_{20}$ in at%. The metallic layers and MgO interlayer were produced by DC and RF magnetron sputtering, respectively. All the samples were post-annealed at 300°C for 1 h under a magnetic field of 6 kOe applied perpendicular to the film plane under a base pressure of $10^{-6}$ Torr. We patterned the film into a 5 μm-width and 35 μm-length Hall cross-device using photolithography (Karl Suss MA6) and Ar ion milling. The electrical contact pad made of the Ti/Au bilayer was fabricated using an e-beam evaporator and subsequent lift-off process. We also constructed 4 × 4 μm²-sized ferromagnetic islands at the center of the Hall bar devices for current-induced SOT switching where the other parts except the islands were W/W-X layers.

**Measurements**

Hysteresis loops were measured using a vibrating sample magnetometer (VSM, Microsense EV9). Evaluating the SOT efficiency, we performed the harmonic Hall response method (Supplementary Note 6). We also utilized the two lock-in amplifiers to simultaneously access the first- and second-harmonic responses of the magnetization. We injected 13.7 Hz and 1 mA amplitude ac current during the measurement while changing the external magnetic field from -18000 Oe to 18000 Oe with different azimuthal angles $\varphi$. To prevent the formation of multidomain, we intentionally tilted the polar angle of the sample (~5º). The current-induced SOT switching was measured by the anomalous Hall voltage of the samples. The Hall voltage was detected by 100 μA dc current after applying every current pulse at 0.5 mA steps and a 10 μs width from -15 mA to 15 mA under a constant external field with *x*-direction. The resistivity of the alloyed layer was evaluated using a constant DC supply. Both switching and resistivity were examined using a 4-point electrical property measurement station (MSTECH M7VC). Crystal structures of the W/W-X layer were characterized by grazing incidence (GI) X-ray diffraction measurements with the incident angle ranging from 30º to 80º at 0.02º steps (XRD, Rigaku ATX-G). To examine the atomic distribution after 300ºC annealing, we used a scanning transmission electron microscope (STEM, FEI Double Cs Corrected Titan3 G2 60-300) for energy-dispersive X-ray spectroscopy. For STEM sampling, we adopted a focused ion beam (FIB, FEI Quanta3D) system. We also employed secondary ion mass spectroscopy (SIMS, ION-TOF TOF.SIMS 5) to examine atomic distribution.

**Result and discussion**

**Materials screening through *ab-initio* calculations**

The large $\theta_{SH}$ of β-W is attributed to large spin Hall conductivity as well as high resistivity, where "acceptor" alloying in β-W enhances the spin Hall conductivity[21]. Such "acceptor" alloying can be achieved with Ta and V, having one less valence electron than W. More specifically, introducing Ta or V can elucidate the resonant feature of the spin-orbit splitting of the doubly degenerate band near the Fermi level, where composition plays a critical role. Previous study[21] focused on $x$ = 12.5 at% alloying with Ta alloying, which is now extended to other compositions of Ta and V.

In β-W, in so-called A15, the structure has two symmetrically inequivalent sites: body-centered-cubic (*bcc*) and chain (*c*) sites. Due to the presence of these two inequivalent sites, several possible configurations exist for each *x*, where the popularity of configuration is determined by energetics. The spin Hall conductivity ($\sigma_{SH}$) is estimated by the thermodynamic average (Supplementary Note 1). In particular, for $x$ = 25 at%, there are four possible configurations: *cc*, *bc*, *cc'*, and *bb'*, where *b* and *c* denote either a *bcc* site or a chain site. More specifically, *cc* and *cc'* configurations exist when Ta or V takes the same and different chain sites, respectively. Similarly, *bb'* denotes when two different *bcc* sites are replaced; *bc* is when each of the *bcc* and chain sites is substituted (Fig. S1 in Supplementary Note 1).

Spin Hall conductivity ($\sigma_{SH}$) is theoretically calculated as a function of $x$ (see Fig. 1a and b) as well as ***k***-resolved Berry curvatures of both W-Ta and W-V alloy when $x = 25$ at% (see Fig. 1c and d). Two alloys exhibit quite different behaviors with composition. The $\sigma_{SH}$ for W-Ta decreases rather monotonically with $x$, whereas that in W-V alloy is enhanced when $x = 12.5$~50 at%. The W-Ta alloy favors the *cc'* configuration with a 96% probability; hence other configurations are not considered. On the other hand, for W-V alloy, *bb'* configuration is the highest with a 60.5% probability, while *bc* configuration has a 35% probability. Probabilities are estimated by the Boltzmann factors, taking the relative total energies into account at room temperature.

The spin Hall conductivity of the *bc* configuration reaches as high as $-1.98 \times 10^3$ S/cm, which is a 141% enhancement over $-0.82 \times 10^3$ S/cm of β-W (see Fig. S1e in Supplementary Note 1). The enhancement is attributed to stronger symmetry breaking when V takes the *bcc* and chain sites simultaneously. More specifically, the Berry curvatures with opposite signs, a characteristic of β-W with the resonant double degenerate state, no longer cancel out. Instead, lifted degeneracies associated with lowered symmetry result in an increase in spin Hall conductivity, whose effect is most drastic when $x = 25$ at% in the W-V alloy. Based on the calculations, W-V alloy with a composition in the range of 12.5 ~ 50 at% is explored experimentally, as discussed below.

**Spin-orbit torque efficiency estimation employing harmonic Hall method**

Based on the first-principles calculations, we introduce the alloy interfacial layer between W and the CoFeB interface in the W/CoFeB/MgO/Ta structures. Based on our previous study, the W layer could not maintain the β-W phase when it was fully alloyed with 10 at% of Ta[29].

A film stack of β-W 4/$W_{100-x}X_x$ 2/CoFeB 0.9/MgO 1/Ta 2 (layer thicknesses in nm) was fabricated with alloy components X = Ta or V to prevent the phase transition of W from β to α. Note that the β-phase is a fundamental assumption of the theoretical calculations.

The alloyed layer composition ranged from 0 to 100 at% of X with steps of 10 at%. After annealing at 300ºC for 1 h in a vacuum, most of the film structures showed PMA. However, when V composition exceeded 80 at%, the film lost the PMA feature and turned to in-plane magnetic anisotropy (IMA). In this composition, the value of effective magnetic anisotropy energy ($K_u^{eff}$) diminished. (Supplementary Note 3). When $x = 100$ at%, the magnetic hysteresis loop is not observed, hence the sign of $K_u^{eff}$ cannot be determined. The degradation of magnetic properties when $x > 80$ at% is attributed to the formation of a magnetic dead layer between V and CoFe[30, 31]. Henceforth, the SOT efficiency is discussed for the composition range with PMA.

To assess SOT efficiency, the harmonic Hall measurements are performed[32]. The SOT consists of two orthogonal vector components, i.e., the damping-like (DL) and the field-like (FL) torque. Fig. 2a schematically illustrates the harmonic Hall measurement, where AC current is injected along the x-direction, subsequent to which the Hall voltage ($V_H$) along the y-direction is measured. During $V_H$ measurement, sweeping the external in-plane magnetic field parallel (perpendicular) to the current yields the trace of DL-SOT (FL-SOT).

Fig. 2b and c show the DL-SOT ($\xi_{DL}$) and FL-SOT ($\xi_{FL}$) efficiency of the W/$W_{100-x}X_x$/CoFeB/MgO/Ta (X= Ta or V) structures for alloy components X, respectively. As shown in Fig. 2b and c, the changes in $\xi_{DL}$ and $\xi_{FL}$ with respect to X are quite different. When X = Ta, as the content of Ta increases, $\xi_{DL}$ gradually decreases from -0.35 ± 0.002 for 6 nm of W to -0.06 ± 0.06 for W/Ta bilayer. Notably, the value of $\xi_{DL}$ for W/Ta is similar to the highly resistive β-Ta[2,33]. This implies that the spin current generated in W does not efficiently penetrate the W-Ta interfacial layer and further intensified with Ta composition. As a result, in the extreme case of a bilayer NM structure, most of the DL torque is due to the spin current generated in the Ta. Furthermore, it is consistent with the previous studies on the spin diffusion length of Ta, which is 2 to 3 nm[34,35].

In the case of V, however, once V is incorporated at the interface of W and CoFeB, the $\xi_{DL}$ increases with a maximum value of about -0.49 ± 0.05 when $x$ = 20 at%. This large value is compatible with a W-based SOT tunnel junction[19,20], whose origin is discussed later.

On the other hand, $\xi_{FL}$ does not show any distinct compositional dependence, as depicted in Fig. 2c. In both Ta and V cases, $\xi_{FL} \approx$ -0.10 ~ -0.20 implies that the potential gradient at the interface (i.e., Rashba effect) has a small correlation with the inserted layer. In addition, in the thin FM layer, namely, layered structure with PMA, the $\xi_{FL}$ is dependent on the magnetic anisotropy energy and temperature[36]. Moreover, the PMA in the NM/CoFeB/MgO structure mainly originated from the CoFeB/MgO interfacial anisotropy energy, and the portion of NM/CoFeB is relatively small. Thus, the insensitive behavior of $\xi_{FL}$ in our case is in line with the previous report because we only modified the W/CoFeB interface.

In accordance with the *ab-initio* calculations, the spin-Hall conductivity is estimated by adopting the relationship between resistivity and the SHA (i.e., $\theta_{SH} = \sigma_{SH}\rho_{xx} + b$ )[37], where $\sigma_{SH}$ consists of the intrinsic and side-jump contributions and $\rho_{xx}$ is the longitudinal resistivity; $b$ represents the skew-scattering contribution. Presumably, parameter $b$ is negligible because of the high impurity content in our system. Hence, the spin-Hall conductivity can be expressed by the ratio of $\theta_{SH}$ to $\rho_{xx}$. $\rho_{xx}$ of each device is determined by employing the parallel resistance model assuming $\rho_{xx}$ = 170 μΩ cm for CoFeB (Supplementary Note 4). Fig. 2d shows the spin-Hall conductivity of the alloy inserted structures.

**Assessment of spin-orbit torque switching**

Next, we examine the current-induced SOT switching properties in the patterned devices. We perform the current sweep from -15 mA to 15 mA. The current pulse width is 10 μs and 0.5 mA steps while varying the external magnetic field strength from -100 to 100 Oe in the x-direction. An additional fabrication process has been utilized to make an island pattern on the devices to test the current-induced SOT switching (Methods). Fig. 3a shows the switching loops for the 20 at% V alloy inserted device, which has the highest $\xi_{DL}$.

As shown in Fig. 3a, the sample clearly exhibits an external field dependence. When we change the polarity of the external field, the switching polarity also changes. The stronger the magnetic field we apply, the smaller the switching current we observe. Furthermore, we examine the magnetization switching in all patterned devices used in harmonics measurement. As depicted in Fig. 3b, we are able to successfully manipulate the magnetization direction by applying an in-plane current pulse and a constant external field, 100 Oe. According to ref.[38], the switching current density ($J_{SW}$) and the $\xi_{DL}$, namely, $\theta_{SH}$, are inversely proportional:

$$J_{SW} = \frac{2e}{\hbar \theta_{SH}} M_S t_{FM} \left( \frac{H_K^{eff}}{2} - \frac{H_{ext}}{\sqrt{2}} \right) \quad (1)$$

where $t_{FM}$ represents the thickness of the FM layer; $H_K^{eff}$ and $H_{ext}$ are the anisotropy field and the external field, respectively. However, according to Figs. 2b and 3b, $\theta_{SH}$ and the switching current density do not show an inverse proportionality. Since we use a 4 μm × 4 μm sized island pattern to test the SOT switching, the size of the island is too large to maintain the single domain state. Therefore, we attribute such discrepancy to the existence of domains and domain walls[39,40]. In the next section, we test whether SOT switching entails the domain wall motion and double-check the large SOT efficiency by utilizing domain wall depinning and the propagation model for 20 at% of the V-incorporated sample.

**Domain wall depinning and propagation model**

To double-check such a high SOT efficiency in our 20 at% V alloy sample, we apply domain wall depinning and the propagation model[22]. First, we determine whether SOT switching occurs through domain wall propagation or single domain reversal when we inject current. We define the $H_p(\theta)$ as an external switching field at the polar angle $\theta$. Fig. 4a shows the ratio of the $H_p(\theta)/H_p(0)$ exactly following the $1/\cos\theta$ dependence, which confirms that in our Hall cross device, the magnetization switching encompasses the domain wall mediated process.

According to the SOT evaluating process[22], we also estimate the measuring time ($t_m$) dependence of $H_p$, as shown in Fig. 4b. We discover that the intrinsic coercive field ($H_{c0}$) is 393.03 Oe. Additionally, we determine a Joule heating coefficient ($\kappa$) by utilizing two different channel resistance ($R_{xx}$) measurements. One is the temperature dependence of $R_{xx}$ by heating the device externally, the other is the current dependence of the $R_{xx}$ from the relationship $T(I) \approx T_0 + \kappa I^2$, where $T_0$ is room temperature, and $I$ is the amplitude of the input current.

When we apply a current of $I = 4.22$ mA, the temperature of the device increases by 157 K with respect to room temperature, where $\kappa$ is determined to be 0.02485 K/mA$^2$. Using these parameters, we converted the switching state diagram (Fig. 4c) to the DL effective field normalized by input current density (Fig. 4d). The detailed conversion process is described in Note 7 of Supplementary Information.

The DL effective field saturates at around 24.7 ± 2.33 Oe cm$^2$/MA when the external field is above 120 Oe. This is compatible with the effective DL field obtained by the harmonic Hall method, as indicated by the red dotted line in Fig. 4d. From the fact that two different measurement schemes show similar magnitudes of DL-SOT effective field, we conclude that the enhancement of SOT results from the 20 at% V incorporation in the film structure.

Furthermore, we check the *1/cos θ* dependence of all devices and confirm that in our devices, SOT switching occurs in the domain wall-mediated process. From this perspective, SOT switching efficiency, considering the domain wall depinning field, is investigated (Supplementary Note 5). We observe that the enhancement ratio of SOT switching efficiency between pristine W and the 20 at% V incorporated sample is qualitatively consistent with the enhancement ratio of $\xi_{DL}$ (Supplementary Note 5). Therefore, increasing the switching current per the increase in V content in Fig. 3b does not consider the effect of the domain wall depinning field. When the domain wall depinning field during magnetization switching is considered, the W-V alloy shows greater SOT switching efficiency than does the W single layer.

**Enhancement in SOT efficiency in W-V alloys**

We analyzed the microstructure of the films to elucidate the increase in the SOT efficiency of the W-V alloy. By adopting the grazing incident X-ray diffraction (GI-XRD), we examine the phase of the W/W-X layer, as shown in Fig. 5a and b. As the layers are ultrathin, the XRD spectra show no peaks corresponding to CoFeB, MgO, and Ta. The W-Ta alloy case retains the β-W phase up to 60 at% of Ta composition. In the case of the W-V alloy, as illustrated in Fig. 5b, the characteristic peak for the α-W appears when 100 at% of V is incorporated into the interfacial layer. The β-W has the A15 structure, which is an energetically unstable phase. In addition, the atomic radii for W, Ta, and V are, respectively, 135, 146, and 135 pm, so the residual stress in the W-Ta alloy promotes the phase transition of the β-W. As a result, we ascribe the phase stability of β-W in the W-V alloy as one of the causes for enhancing the DL effective field.

To examine whether the enhancement of SOT efficiency was due to the atomic interdiffusion or not, we captured the atomic distribution profiles employing STEM. As shown in Fig. 6a and b, we clearly observed the layered structures. Schematic illustrations of the layered structures are placed on the right side of the figures. Also, we note that the interfaces between each layer became clear and sharp when the films underwent the 300°C annealing. We utilized the EDS to allocate the atomic position in the film structure. In order to secure the EDS intensity, the thickness of the sample prepared for high-resolution (HR) TEM measurement was too thin (20-30 nm), and thus a relatively thick specimen (80-100 nm) was prepared using a focused ion beam (FIB) system. We allocated the atomic position in the film structure through the EDS, as presented in Fig. 6c and d. Note that the high-angle annular dark-field (HAADF) images appear blurred because of the TEM sampling thickness. Also, we have performed the EDS for the samples used in the HR-TEM measurement (Note 8 of Supplementary Information). According to Fig. 6c and d, V atoms were located beneath the CoFe layer regardless of heat treatment. We confirmed that the V's elemental profile, observed in the $SiO_2$/W and Ta layers, is an EDS measurement artifact (Note 8 of Supplementary Information). Further, we adopted SIMS and observed that V's atomic interdiffusion is negligibly small (Note 9 of Supplementary Information). Therefore, we conclude that the improvement of the SOT efficiency in heterostructures incorporating W-V interfacial layers genuinely originated from the W-V interfacial layer between W and CoFeB.

Moreover, when the W-V alloy is introduced, the assumption of the theoretical calculation[21] that the β-W should maintain its crystal structure when it mixes with another element with one less valence band of electrons is fulfilled. We experimentally calculate the maximum SHC value -(2.77 ± 0.31) × $10^3$ S/cm at 20 at% of V, as shown in Fig. 2d. It is reasonably consistent with our first-principle calculation results, which have a maximum value of -1.98 × $10^3$ S/cm at 25 at% of V as shown in Fig. 1b. The introduction of alloy layer between W and CoFeB layers enlarges the charge-to-spin conversion efficiency of W. Both experiment and theory exhibit the same trend quantitatively with respect to the composition of the "acceptor". We think it is worthwhile to investigate the bulk spin Hall properties of standalone W-V layers for further study.

**Conclusion**

A combined study of theory and experiment is presented for semiconductor-process friendly W/W-X/CoFeB/MgO/Ta (X=Ta or V). A series of ab initio calculations are employed to search for a W-X alloy with large spin Hall conductivity. Subsequent fabrication and measurement are performed by experimentally compiling composition-dependencies of SOT efficiency. In particular, a significant enhancement of the DL-SOT efficiency ($\xi_{DL}$) is found for 20% V alloying over the pristine β-W by 40%. This significant enhancement of $\xi_{DL}$ is confirmed by two different measurement schemes, harmonic measurement as well as domain wall depinning and propagation model. Through the XRD and TEM analyses, we confirmed that the enhancement in SOT efficiency was attributed to the phase stability of β-W in the W-V interfacial alloyed structure and originated from the W-V alloy layer, predicted by the theoretical calculations. In summary, the combined study of theory and experiment sheds light on engineering opportunities to develop materials exhibiting high charge-to-spin conversion efficiency.

**Acknowledgements** We thank Yun Jung Jang of the Korea Institute of Science and Technology (KIST) for helping us in ToF-SIMS experiments. This research is supported by the National Research Foundation (NRF) of Korea grant funded by the Ministry of Science and ICT (2015M3D1A1070465, 2020M3F3A2A01082591) and in part by the Samsung Electronics Co., Ltd. (IO201211-08104-01). The work at the University of Ulsan is supported by the Basic Research Lab Program through the NRF of Korea funded by the Ministry of Science and ICT (2018R1A4A1020696).

**Author information**

These authors contributed equally: Gyu Won Kim, Do Duc Cuong

**Affiliations**

**Department of Materials Science and Engineering, Korea University, Seoul 02841, Korea**

Gyu Won Kim, Yong Jin Kim, In Ho Cha, Taehyun Kim, Min Hyeok Lee

**Department of Physics, University of Ulsan, Ulsan 44610, Korea**

Do Duc Cuong, Soon Cheol Hong, Sonny H. Rhim

**Center for Spintronics, Korea Institute of Science and Technology, Seoul, 02792, Korea**

OukJae Lee

**Seoul Center, Korea Basic Science Institute, Seoul, 02841, Korea**

Hionsuck Baik

**Contributions**

GWK and YKK conceived and designed the experiments. DDC, SCH, and SHR performed the first-principles calculations. GWK, TK, and MHL fabricated film structures and devices. GWK,

YJK, and IHC conducted the SOT efficiency evaluation and analyzed the microstructural transition of β-W. GWK conducted domain wall propagation experiments with the help of OJL. HB performed a TEM experiment and analyzed the atomic distribution. GWK and DDC wrote the manuscript after discussion with all the authors. YKK supervised the entire project.

**Conflict of interest**

The authors declare that they have no conflict of interest.

**Publisher's note**

Springer Nature remains neutral with regards to jurisdictional claims in published maps and institutional affiliations.

**Supplementary information** is available for this paper at https://doi.org/

**Figures with captions**

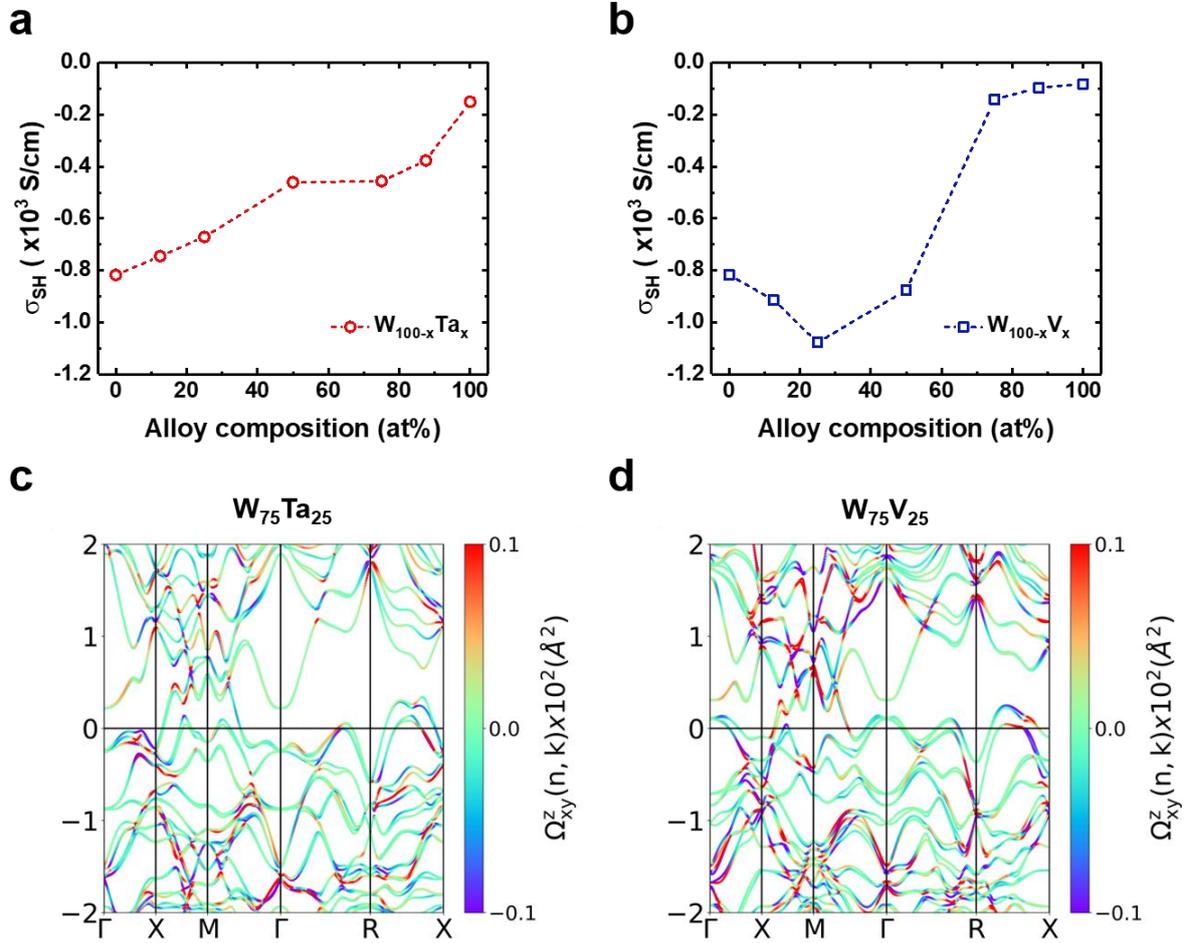

**Fig. 1 Theoretical calculations of spin Hall conductivity. a,b,** Spin Hall conductivities of $W_{100-x}Ta_x$ **(a)** and $W_{100-x}V_x$ **(b)**. k-resolved Berry curvature of $W_{100-x}Ta_x$ **(c)** and $W_{100-x}V_x$ **(d)** alloy where x = 25, where two Ta or V takes place c-c' and b-c configurations, respectively (Fig. S2 in Supplementary Note 2).

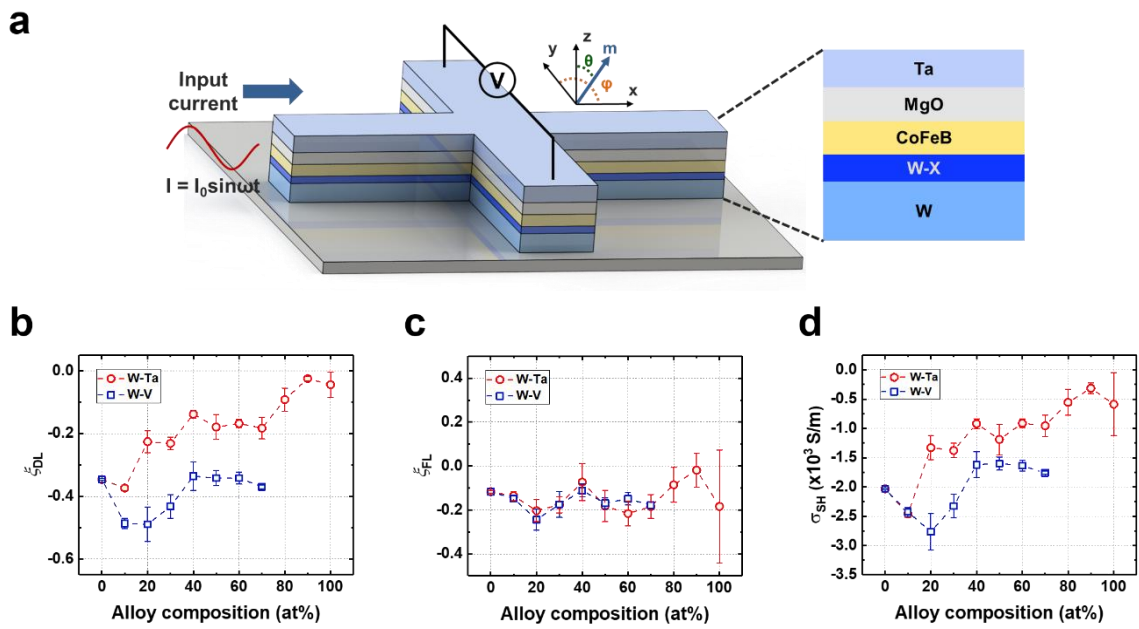

**Fig. 2 Spin-orbit torque measurement. a,** Schematic description of the device structure and stack of the films. We denote the coordinate axis above the device structure. **b, c,** DL, and FL SOT efficiencies, respectively. **d,** Experimentally estimated spin-Hall conductivity. Error bars in the figures represent the standard error.

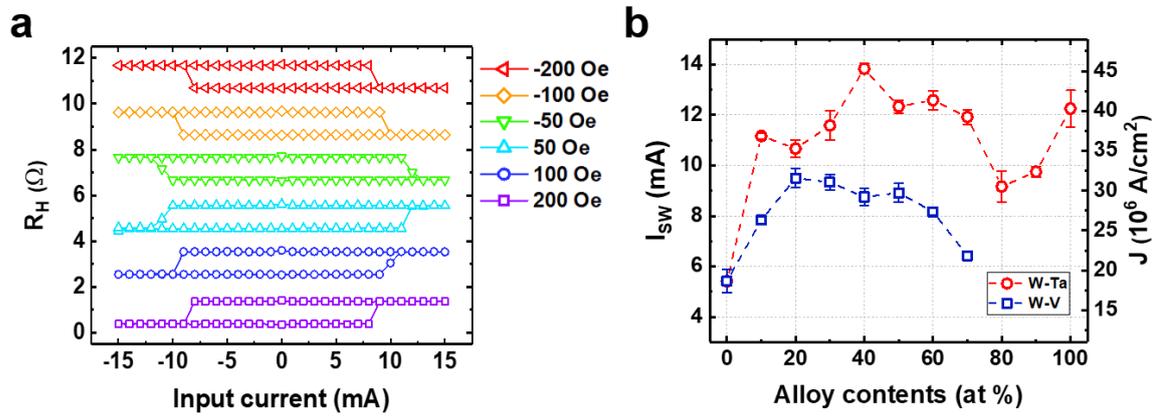

**Fig. 3 In-plane pulsed current-induced spin-orbit torque switching. a,** SOT switching curves for different external field strengths for the 20 at% of V incorporated device. **b,** SOT switching current density for the whole PMA samples as a function of alloy composition. Under constant 100 Oe external magnetic field along the x-direction.

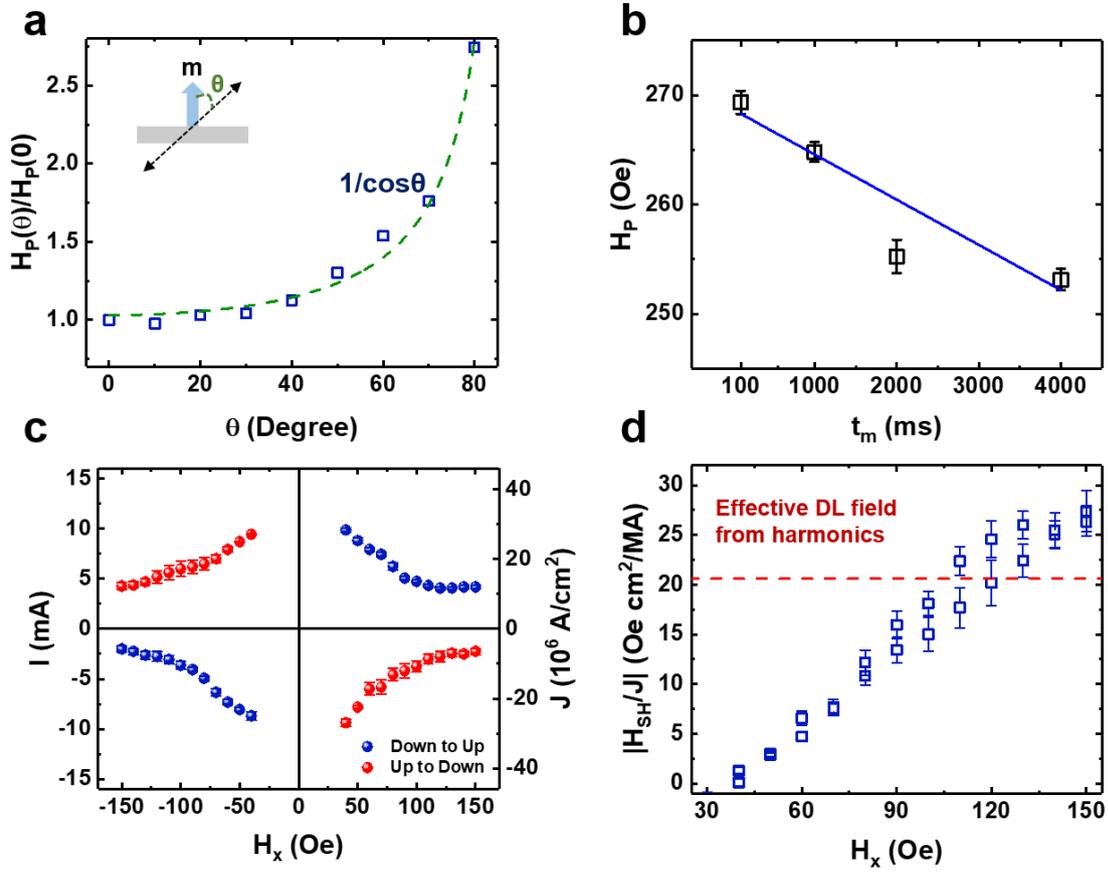

**Fig. 4 Domain wall depinning and propagation switching model in the 20 at% of V alloy inserted structure. a,** Polar angle dependence of switching field. The green dotted line represents the fitted line for *1/cos θ*. **b,** Measuring time ($t_m$)-dependent behavior of the switching field. Error bars represent the standard deviation from the 35 times independent measurements. **c,** Switching state diagram as a function of the external field in the x-direction. **d,** Effective DL field evaluated from the switching state diagram. The red dotted line represents the effective DL field estimated from the harmonic response technique. Error bars in the figures represent the standard error.

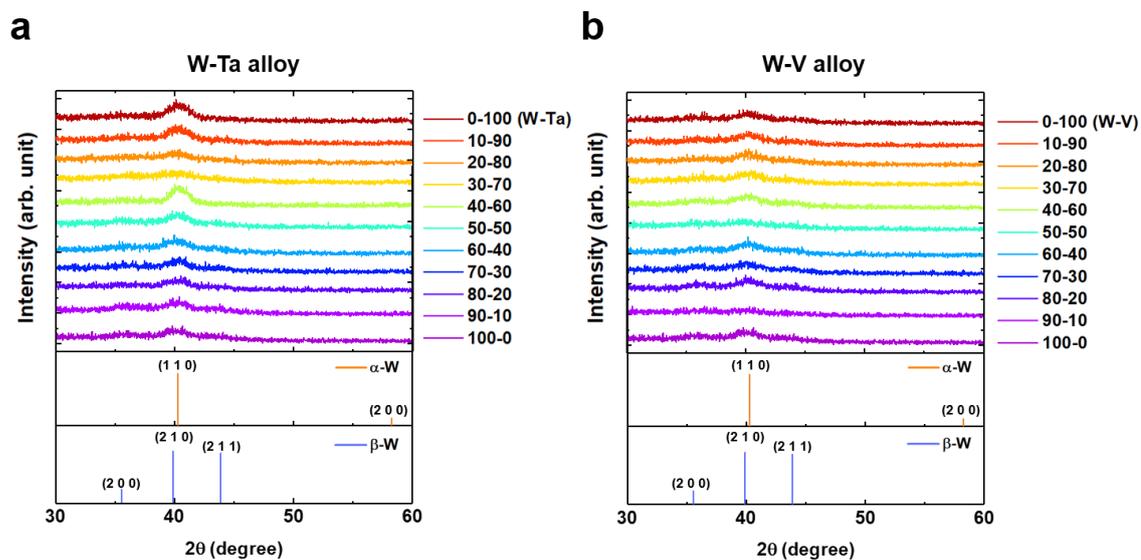

**Fig. 5 Microstructural analysis for alloy inserted structure. a,b,** XRD spectra for the W/W$_{100-x}$X$_x$/CoFeB/MgO/Ta films for varying compositions of inserted alloy film W$_{100-x}$X$_x$ after annealing at 300°C for 1 h, X = Ta **(a)** and X = V **(b)**. Reference peak positions for α (ICDD No. 00-004-0806) and β (ICDD No. 00-047-1319) phase of W are denoted below the spectra by the vertical lines.

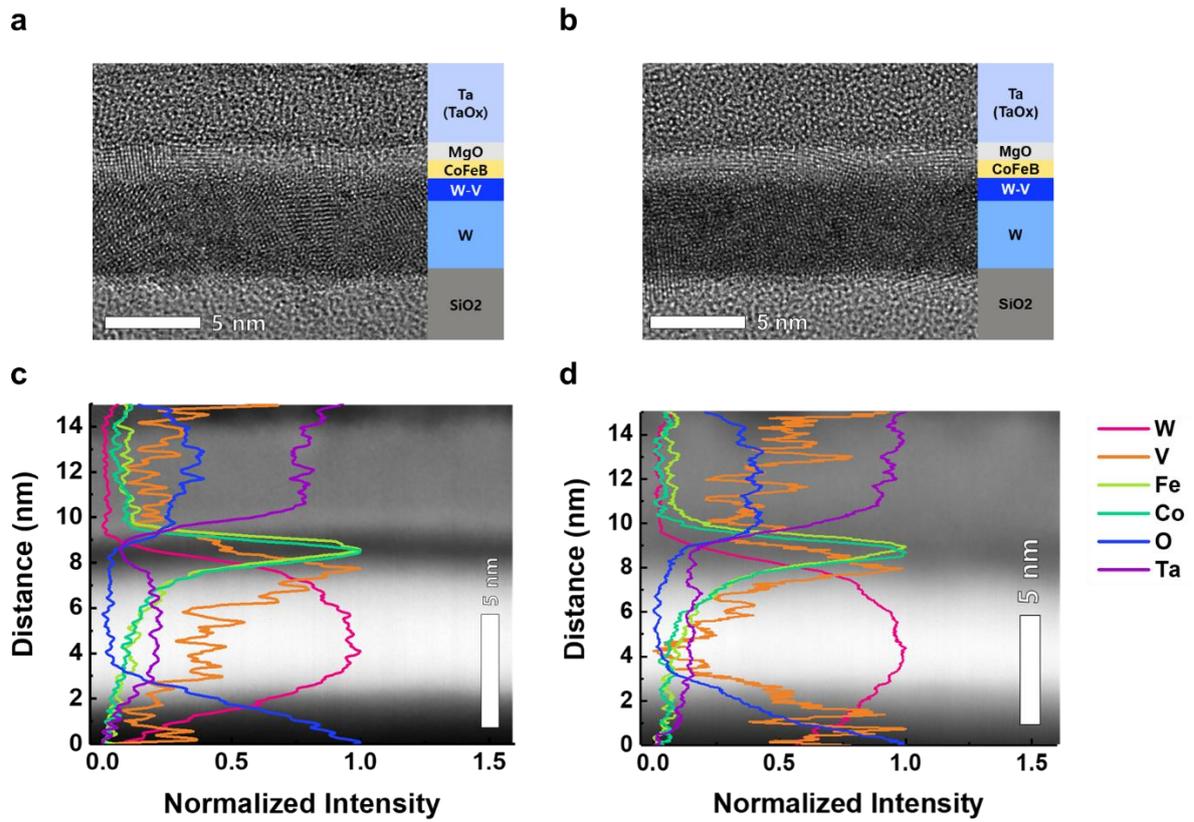

**Fig. 6 Scanning transmission electron microscopy (STEM) data for the W/W-V/CoFeB/MgO/Ta structure. a, b,** HR-TEM image for as-deposited and 300°C annealed sample. **c, d,** HAADF-STEM image and element line profiles for as-deposited and 300°C annealed samples. The composition of the alloy layer is $W_{80}V_{20}$ in the atomic ratio.

**SUPPLEMENTARY INFORMATION**

**Spin-Orbit Torque Engineering in β-W/CoFeB Heterostructures via Ta and V Alloying at Interfaces**

Gyu Won Kim[1†], Do Duc Cuong[2†], Yong Jin Kim[1], In Ho Cha[1], Taehyun Kim[1], Min Hyeok Lee[1], OukJae Lee[3], Hionsuck Baik[4], Soon Cheol Hong[2], Sonny H. Rhim[2*], and Young Keun Kim[1*]

Correspondence: Young Keun Kim (ykim97@korea.ac.kr) or Sonny H. Rhim (sonny@ulsan.ac.kr)

[1]Department of Materials Science and Engineering, Korea University, Seoul 02841, Korea.

[2]Department of Physics, University of Ulsan, Ulsan 44610, Korea.

[3]Center for Spintronics, Korea Institute of Science and Technology, Seoul 02792, Korea.

[4]Seoul Center, Korea Basic Science Institute (KBSI), Seoul, 02841, Korea

A full list of author information is available at the end of the article

[†] These authors contributed equally: Gyu Won Kim, Do Duc Cuong

**Contents**

**Note 1. Thermodynamic average with respect to the atomic configuration**

In our first-principles calculations, at each concentration (*x*), all possible configurations are taken into account. As mentioned briefly in the main text, β-W in A15 structure consists of two types of symmetry sites, *bcc*-and chain-sites. The chain sites form chain-like structures along the x, y, and z directions, a distinct feature of the A15 structure. The case of *x* = 25% is depicted in Fig. S1a-d, where configurations are labeled as *cc*, *bc*, *cc'*, and *bb'*, which denote chain-chain (equal chain), *bcc*-chain, different chains, and two *bcc* sites, respectively. The spin Hall conductivity of each x is calculated for all possible configurations, where the *x* = 25% case is shown in Fig.S1e. $\sigma_{xy}$ of a particular *x* is evaluated by thermodynamic average, i.e., $\sigma_{xy}$ of each particular configuration is weighted by the Boltzmann factor, $exp(-\Delta E/k_B T)$, where $\Delta E = E^\alpha - E_0$, is the relative energy of configuration α with respect to the lowest energy configuration ($E_0$) and room-temperature is taken for the estimate. Relative energetics of each configuration are shown in Fig. S1f: the *cc'* configuration is the most favored for the W-Ta alloy and the *bb'* for the V-W alloy. In the case of W-Ta, *cc'* configuration absolutely dominates over others. In the V-W case, on the other hand, the *bc*- configuration has the highest spin Hall conductivity, but its relative probability is 4.03%. However, *bb'* configuration, with the lowest energy, 60.5 % probability, has $\sigma_{xy}$ of -0.89×10³ S/cm, and *cc'* with a 35.39 % probability has $\sigma_{xy}$ of -1.29×10³ S/cm.

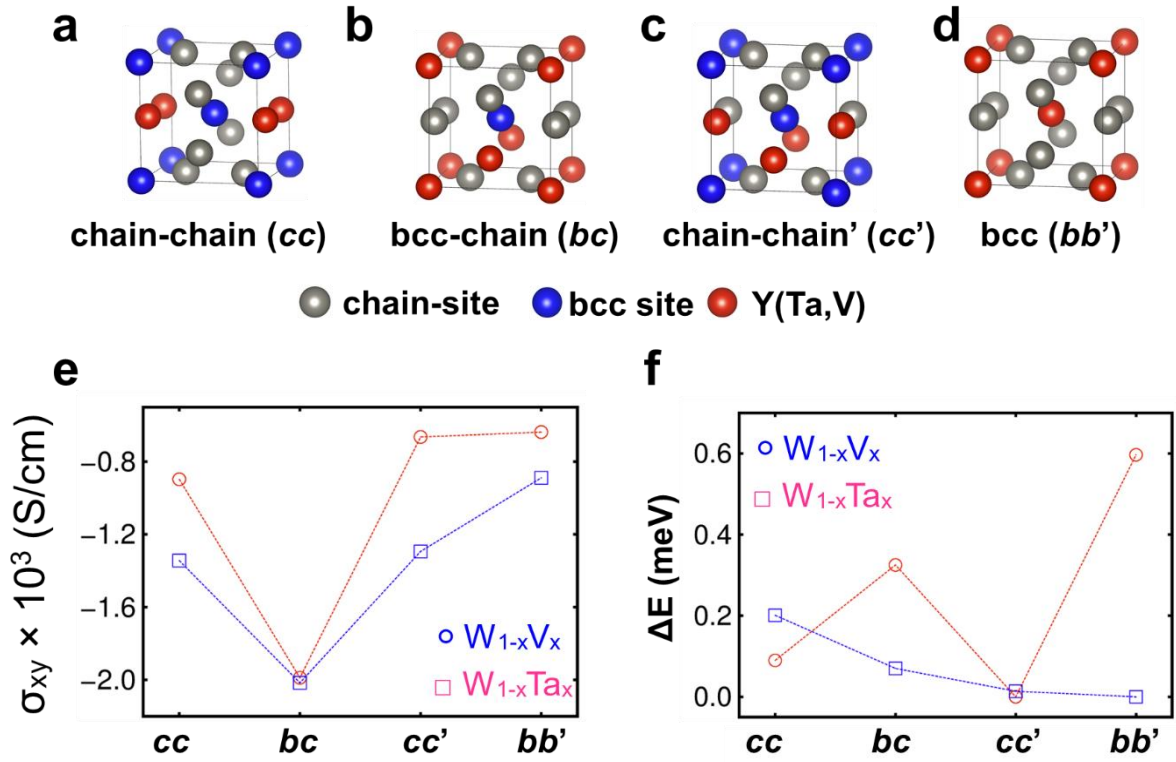

**Fig. S1 Possible configurations for x = 25% in the primitive cell of A15 structure and spin hall conductivities, thermodynamic energy differences for each possible structure.** Ta or V can take place in either chain or bcc sites. We denote when two sites of one chain are taken as chain-chain (*cc*) **(a)**, one at *bcc* the other at chain site as bcc-chain (*bc*) **(b)**, two different chain sites as chain-chain' (*cc'*) **(c)**, and two sites of bcc as bcc-bcc (*bb'*) **(d)**. **e,** Spin Hall conductivity ($\sigma_{xy}$) for four different configurations of Ta-W and V-W alloys at 25% concentration. **f,** Energetics of four configurations, where $\Delta E = E - E_0$ is the relative energy difference with respect to the lowest energy configuration for each alloy ($E_0$).

**Note 2. Anatomy of W-V spin Hall conductivity for x = 12.5 and 25 at%**

Fig. S2 shows $k$ resolved Berry curvature and site-resolved band structure of W-V alloy with $x =12.5$ and concentration of 25 at%. At first sight, two compositions exhibit similar behavior. However, one notes differences. Degeneracy lifting is more drastic in $x = 25$ at% than $x = 12.5$ at%, which is natural considering the lowered crystal symmetry with increased alloy concentration.

Furthermore, near the Fermi energy ($E_F$), band hills become unoccupied in $x = 25$ at% as a result of reduced valence electrons by the introduction of V atoms. While $\Gamma$ points in both concentrations exhibit large Berry curvature, notably, in $x = 25$ at% Berry curvature at the $\Gamma$ point around -1 eV below $E_F$ contributes more, enhancing the overall spin Hall conductivity. To capture the essence, the site-resolved band structure is presented. In $x = 12.5\%$, *bcc*-site contributes more around $E_F$, which is somewhat separate from chain-site contributions. On the other hand, chain-site contribution emerges near $E_F$. Moreover, this emergence chain-site contribution plays a role in stronger degeneracy lifting, which is more evident in the $x = 25\%$ case.

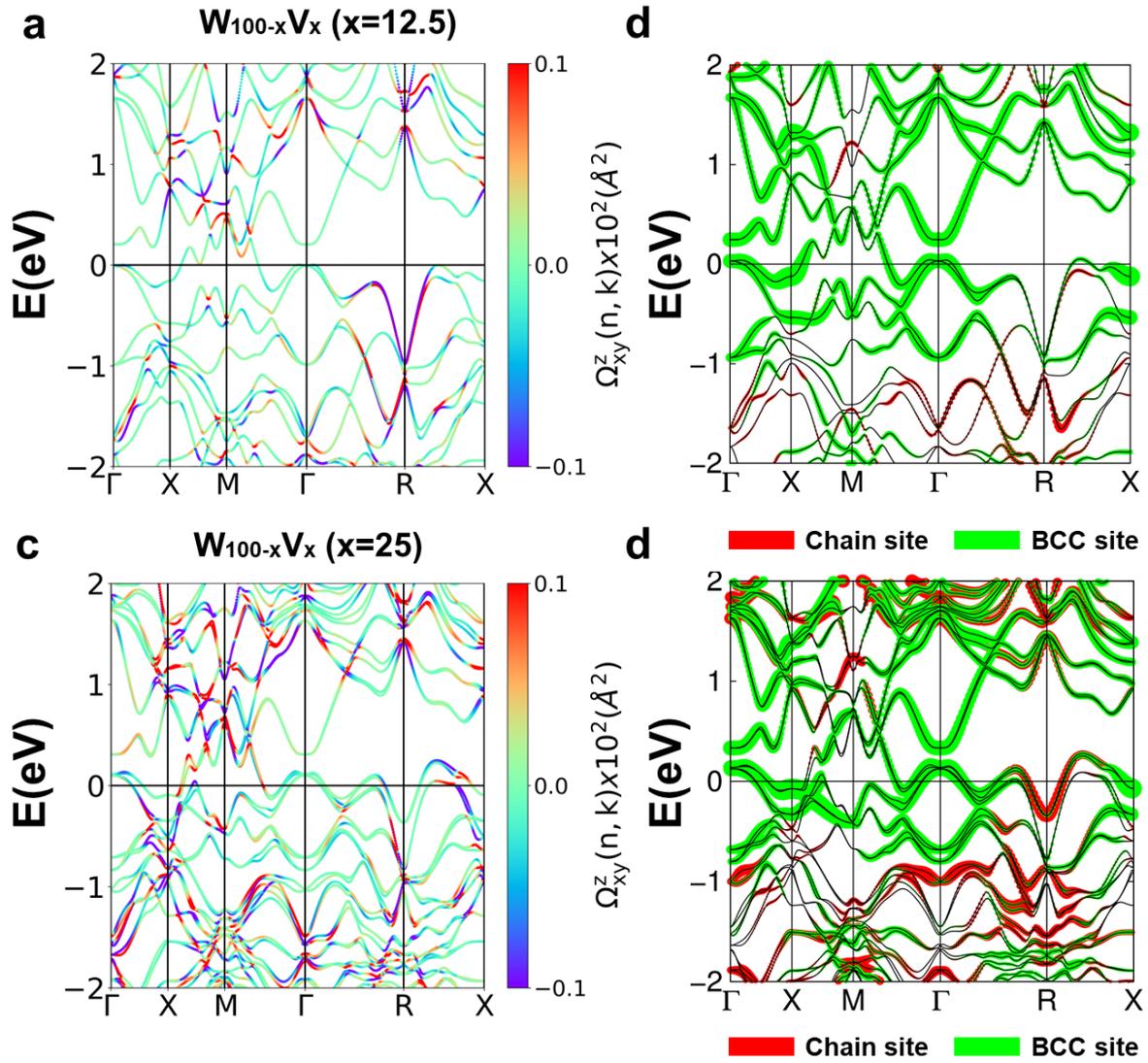

**Fig. S2 Band structure analysis for $W_{1-x}V_x$ alloy for x=12.5% and *x*=25%.** Left and right panels are *k*-resolved Berry curvature **(a, c)** and site-resolved band structure **(b, d)**, where site means either chain or *bcc* sites.

**Note 3. Magnetic anisotropy of the films**

We determined the saturation magnetization and magnetic anisotropy of annealed samples. Table S1 represents the saturation magnetization value for each sample, and Fig.S3 a-d shows the out-of-plane (a, c) and in-plane (b, d) hysteresis loops for W-Ta and W-V alloy. Fig. S3e and f represent the effective magnetic anisotropy energy ($K_u^{eff}$) of the W-Ta and W-V alloys with different compositions, respectively. $K_u^{eff}$ is estimated from the difference of the integrated area between the magnetic easy and hard axis magnetization curves [S1].

According to Fig.S3e and f, for W-Ta alloy, $K_u^{eff}$ is positive for the entire composition indicating strong PMA. For the pristine W, $K_u^{eff}$ = 1.20 × 10⁶ erg/cm³ and for 60 at% Ta $K_u^{eff}$ reaches as high as 2.64 × 10⁶ erg/cm³. However, in the case of V, $K_u^{eff}$ > 0 (PMA) occurs when 0 < x < 80 at%, whose value is 0.14 × 10⁶ erg/cm³ to 1.52 × 10⁶ erg/cm³. When the concentration of V exceeds 90 at%, PMA becomes in-plane magnetic anisotropy. For 100 at% of V, any magnetic hysteresis loops in both out-of-plane and in-plane directions, so we could not determine the $K_u^{eff}$ of the film.

**Table S1. The saturation magnetization ($M_s$) values of samples after 300ºC, 1 h annealing.**

| Alloy composition (W-X, at%) | Saturation magnetization, $M_s$ (emu/cm³) | | | | | | | | | | |
|---|---|---|---|---|---|---|---|---|---|---|---|
| | 100-0 | 90-10 | 80-20 | 70-30 | 60-40 | 50-50 | 40-60 | 30-70 | 20-80 | 10-90 | 0-100 |
| V | 860 | 899 | 867 | 828 | 761 | 774 | 735 | 674 | 633 | 625 | - |
| Ta | 860 | 914 | 867 | 868 | 924 | 828 | 849 | 854 | 655 | 527 | 654 |

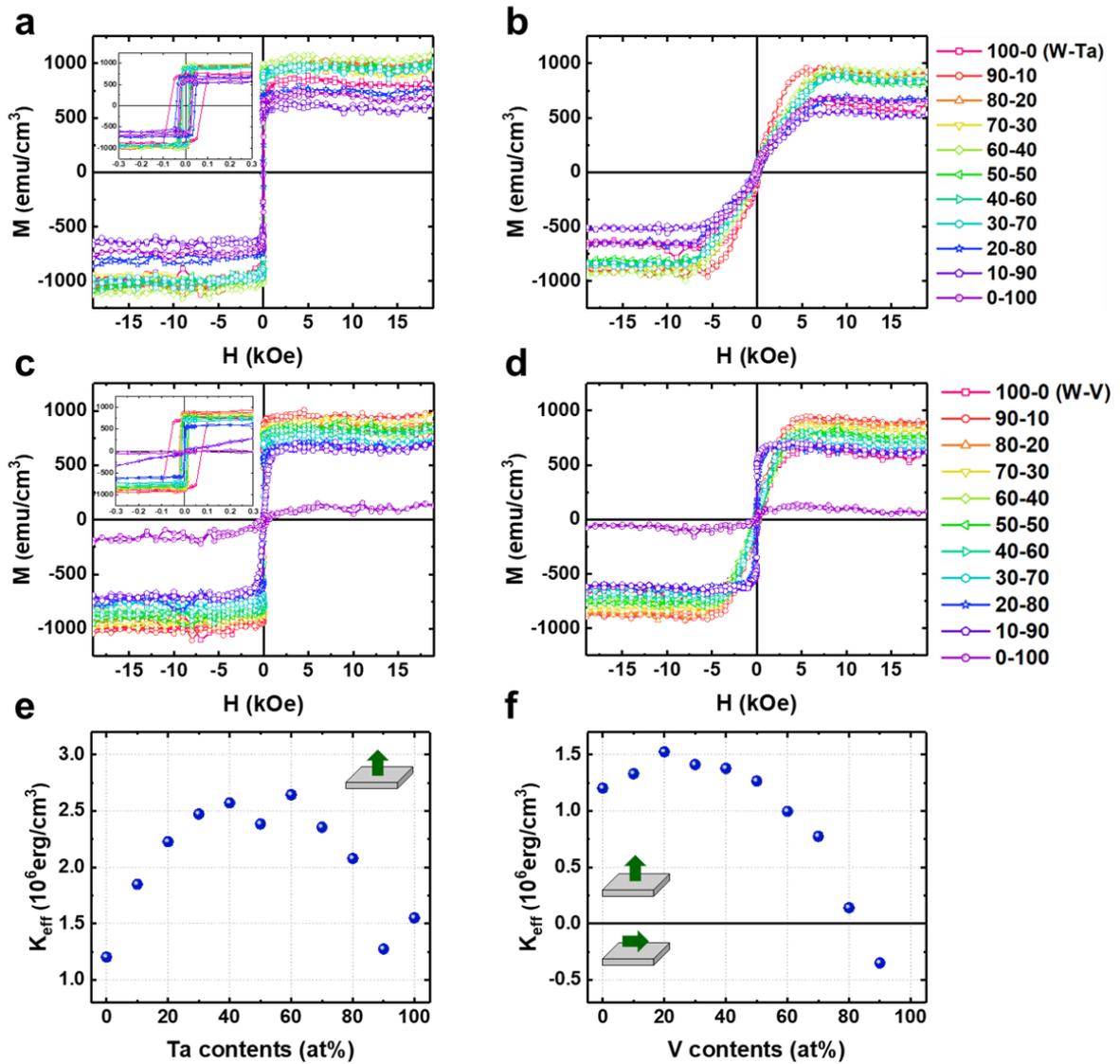

**Fig. S3 Magnetic hysteresis loops and effective magnetic anisotropy energy for W-Ta and W-V alloy systems. a-d,** Magnetic hysteresis loops measured at the out-of-plane (**a, c**) and in-plane (**b, d**) directions. **e, f,** Effective magnetic anisotropy energy of interfacial W-Ta (**e**) and W-V (**f**) alloy case. Insets in the **a** and **c** are the enlarged loops ranging from -0.3 to 0.3 kOe.

**Note 4. Characterization of alloy resistivity**

Fig. S4a shows the optical microscope image for the Hall bar structure, which is used for evaluating the resistivity of the W/W-X layer. Fig. S4b indicates that the W-V alloy system has a higher electrical resistivity than that of W-Ta. According to the supplementary ref. [S4-S6] the $\theta_{SH}$ is proportional to the resistivity of the NM layer. Although this model cannot explain the peak position of the $\theta_{SH}$ (e.g., W-V 20 at%), we can demonstrate that the overall trend of $\theta_{SH}$ is proportional to the electrical resistivity of the NM (or alloyed NM) layer.

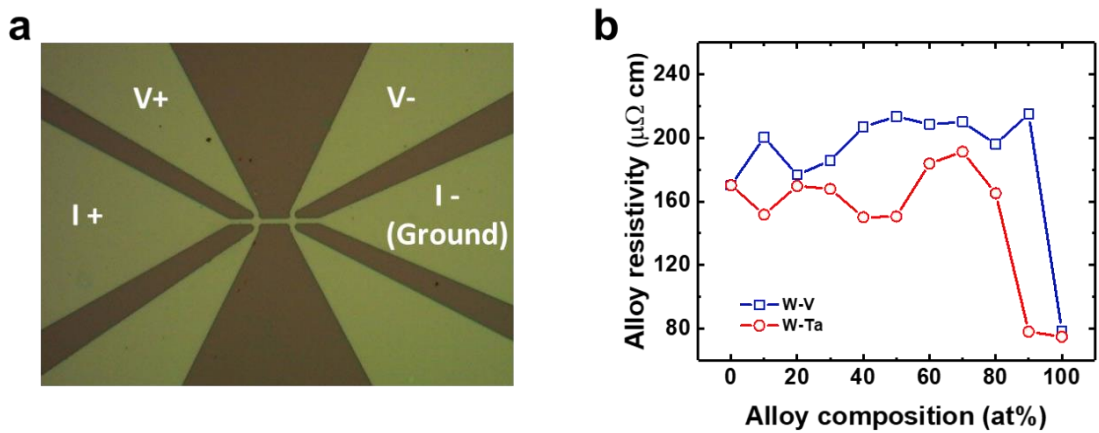

**Fig. S4 Electrical resistivity measurement. a,** Optical microscope image for Hall bar device, which is used in the resistivity measurement. The distance between the voltage lines is 30 μm, and the width is 5 μm. **b.** Electrical resistivity of the W/W-X layer. We adopt a parallel resistance model and assume that areas above the MgO layer did not contribute to the electrical conduction due to oxidation. Here, we use the resistivity of the CoFeB as 170 μΩ cm [S2, S3]. Error bars are smaller than the symbols in the figure.

**Note 5. Domain wall depinning and propagation**

We examine the polar angle dependence of the switching field for the whole composition range to confirm that SOT switching behavior in our devices is a domain wall mediated process. The *1/cos θ* dependence [S7] of the switching field is observed for all samples, and the actual depinning field of every device is denoted in Fig. S5a. Also, we plot the SOT switching efficiency for the W-V series, including the domain wall depinning field (Fig. S5b). According to Fig. S5b, SOT switching efficiency increases when the V is alloyed with W until 30 at%. The enhancement ratio ($\eta$) of SOT (first term) and SOT switching efficiency (second term) between pristine W and V 20 at% sample is calculated using the equation as follows:

$$\eta = \frac{(\xi_{DL})_{W-V}}{(\xi_{DL})_W} = \frac{(I_{SW}/H_P)_W}{(I_{SW}/H_P)_{W-V}}$$

According to the above equation, we calculate the enhancement ratio of SOT as 0.49/0.35 = 1.40 and SOT switching efficiency as 0.042/0.030 = 1.40. This result is confirmed when SOT switching occurs with domain wall motion; one should consider the domain wall depinning field to compare SOT switching current and SOT switching efficiency.

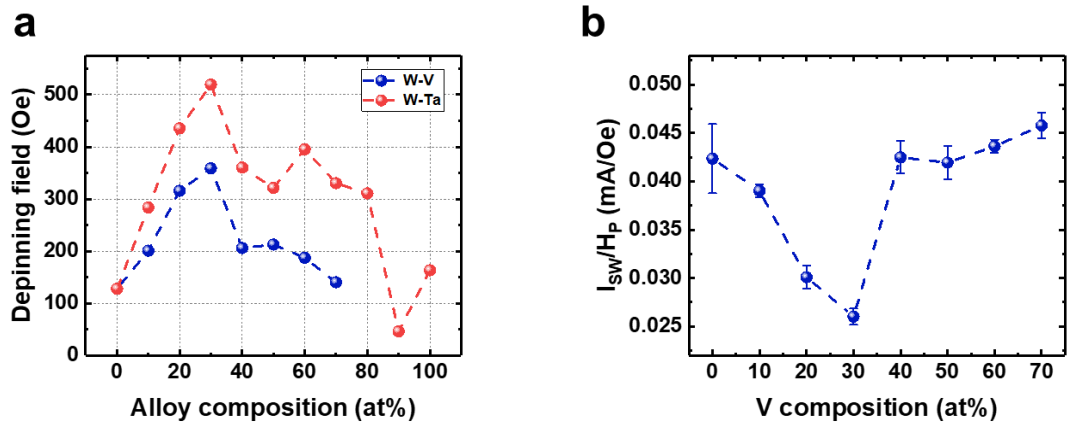

**Fig. S5 Domain wall mediated SOT switching. a,** Domain wall depinning field for interfacial W-Ta, and W-V structure. **b,** SOT switching efficiency for the W-V system considering domain wall depinning field.

**Note 6. Details of harmonic Hall method**

Harmonic responses of the magnetization with polar ($\theta$) and azimuthal ($\varphi$) angular dependence is expressed as:

$$R_H^{1w} = R_{AHE} \cos\theta + R_{PHE} \sin^2\theta \sin 2\varphi \tag{1}$$

$$R_H^{2w} = [R_{AHE} - 2R_{PHE} \cos\theta \sin 2\varphi] \frac{d\cos\theta}{dB_I} \cdot B_I + R_{PHE} \sin^2\theta \frac{d\sin 2\varphi}{dB_I} \cdot B_I,$$

$$+R_T \sin\theta \cos\varphi \tag{2}$$

where $R_{AHE}$, $R_{PHE}$, and $R_T$ represent, respectively, anomalous, planar, and thermo-electric Hall resistance, and $B_I$ is the current-induced field, including the DL, FL effective field, and the Oersted field. When we set the external field larger than the anisotropy field of the device with $\theta = \pi/2$ direction, Eq. (2) is simplified as [S8]:

$$R_H^{2w} = \left[ \left(R_{AHE} \frac{B_{DL}}{B_{eff}} + R_T\right) \cos\varphi + 2R_{PHE}(2\cos^3\varphi - \cos\varphi) \frac{B_{FL} + B_{Oe}}{B_{ext}} \right] \tag{3}$$

$B_{DL}$, $B_{FL}$, $B_{eff}$, $B_{Oe}$, and $B_{ext}$ denote the SOT induced DL and FL effective field, the effective anisotropy field, the Oersted field, and the applied external field, respectively. Since we applied sufficiently large $B_{ext}$, two or three times larger than the anisotropy field of the device, we assume the magnetization is completely aligned to the $B_{ext}$ direction. Accordingly, in this configuration, the thermo-electric contribution to $R_H^{2w}$ is constant. Also, if we rotate the device to $\varphi = \pi/4$, $3\pi/4$, $5\pi/4$, and $7\pi/4$, the second term in Eq. (3) vanishes. Therefore, the thermoelectric signal and $B_{DL}$ are separated in the first term, and $B_{FL} + B_{Oe}$ is estimated from the subtraction of the $\cos\varphi$ dependence in Eq. (3). Also, we calculate the SOT efficiency using Eq. (4):

$$\xi_{DL(FL)} = \frac{2e}{\hbar} M_S t_{FM} \frac{B_{DL(FL)}}{J_e} \tag{4}$$

Here, $e$, $\hbar$, $M_S$, $t_{FM}$, and $J_e$ represent the elementary charge of the electron, the Planck constant, the saturation magnetization, the thickness of the ferromagnetic layer, and the input current density, respectively. We recall, in general, that DL efficiency ($\xi_{DL}$) is treated as the $\theta_{SH}$ of HM.

Since $V_H$ is composed of the product between the injected ac current ($I=I_0 \sin \omega t$) and device resistance, it can be denoted as Hall resistance ($R_H$) simply divided by $I_0$. The $R_H$ has an in-phase first harmonic ($R_H^{1w}$) and out-of-phase second harmonic ($R_H^{2w}$) signals.

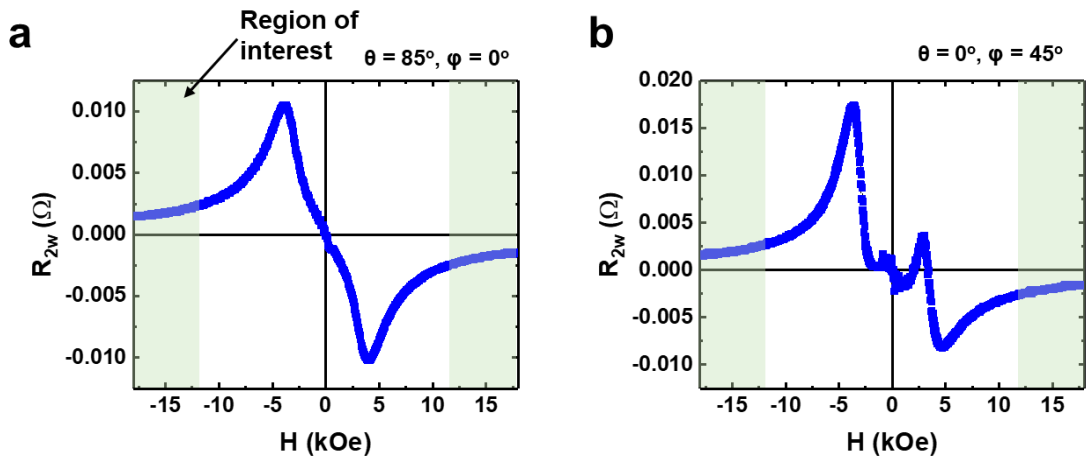

**Fig. S6 Second harmonic data with different polar and azimuthal angle. a, b,** Second harmonic data for W/CoFeB/MgO/Ta structure, measured at $\theta = 85°$, $\varphi = 0°$ **(a)**, and $\theta = 0°$, $\varphi = 45°$ **(b).**

**Note 7. The detailed conversion process from switching current density to DL-SOT efficiency**

When magnetization switching entails the domain wall depinning and propagation process, it can be written as follows:

$$H_c = H_{c0}\left\{1 - \left[\frac{k_B T}{E_a}\ln\left(\frac{f_0 t_m}{\ln 2}\right)\right]^1\right\} \quad (1)$$

where $H_c$, $H_{c0}$, $k_B$, $T$, $E_a$, $f_0$, and $t_m$ represent switching field, depinning field in the absence of thermal fluctuations, Boltzmann constant, temperature, activation energy barrier for depinning, characteristic fluctuation attempt frequency (10 GHz in this case), and sampling time, respectively.

Since thermal stability ($\Delta$) is expressed as $\Delta = E_a / k_B T$, the Eq. (1) can be converted to a current dependent form as follow:

$$\Delta = \left(\frac{E_a(I)}{k_B T(1+\kappa I^2)}\right)\left(1 - \frac{(H_z + H_s(I))}{H_{c0}(I)}\right)^1 \quad (2)$$

In general, thermal stability treat as constant value and if we define $h(I) = \Delta/\Delta_0(I)$, $h_z(I) = H_z/H_{c0}(I)$, $h_s(I) = H_s/H_{c0}(I)$, the Eq. (2) is simplified as Eq. (3).

$$h(I)(1 + \kappa I^2) = \left(1 - h_z(I) - h_s(I)\right)^1 \quad (3)$$

Here, $H_z$, $H_s$, and $\kappa$ represent the applied external field along the z-direction, current-induced spin-orbit torque field, Joule heating coefficient, respectively. We determined $h(I)$ using the ratio of the switching field ($H_c$) and the depinning field in the absence of thermal fluctuation ($H_{c0}$). The value of $H_c$ and $H_{c0}$ was evaluated by Fig. S7a and Fig. 4b, respectively.

$$h(I) = 1 - \left(\frac{H_c}{H_{c0}}\right) = 1 - \left(\frac{274.82}{393.03}\right) = 0.30078\ldots$$

Joule heating coefficient ($\kappa$) was estimated by two different measurements, which were (i) $R_{xx}$ vs. input current density and (ii) $R_{xx}$ vs. Temperature. Data from both measurements were shown in Fig. S7b and c. Here, Rxx is the longitudinal resistivity.

According to Fig. S7b and c, when we applied $1.18 \times 10^7$ A/cm$^2$ (4.22 mA) to the device, the resistivity had the same value at 453 K (1503.5 Ω). Since we started measurement (ii) at 296 K, we assigned the $\kappa$ as 0.02485 K/mA$^2$.

When we evaluated the switching current density, we only applied the external field along the x-direction, so the $h_z(I)$ in Eq. (3) goes to zero. The Eq. (3) could then be simplified and expressed as a current-induced spin-orbit torque field as a function of current amplitude in Eq. (4).

$$h_s(I) = 1 - h(I)(1 + \kappa I^2) \tag{4}$$

Using Eq. (4), we converted the SOT switching current to SOT effective field, shown in Fig. 4c and d.

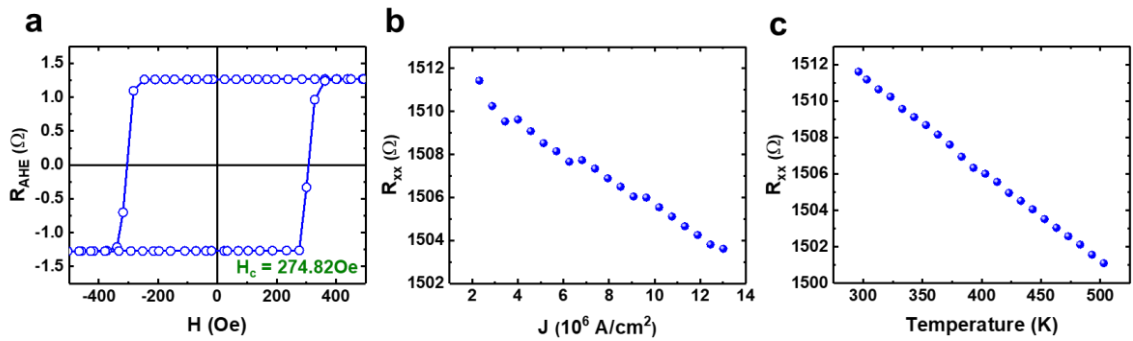

**Fig. S7 Detailed data for estimating SOT efficiency using the domain wall depinning model. a,** field-dependent anomalous Hall voltage **b,** longitudinal resistance as a function of current density, and **c,** longitudinal resistance as a function of sample temperature.

**Note 8. Energy-dispersive X-ray spectroscopy and elemental profiles for the V**

We conducted the HAADF-STEM and EDS elemental scanning on the same samples used for HR-TEM measurement. As presented in Fig. S8a and b, while we obtained a clear HAADF-STEM image for both samples but could not allocate the position of the V because of low intensity. Thus, we prepared other samples which were suitable for EDS elemental line scanning.

Also, to confirm whether the EDS line profiles of V presented in the main text Fig. 6c and d were artifacts of measurement or not, we examined the EDS spectra for the selected area of HAADF-STEM images. We have checked the 3-different regions of as-deposited and 300°C annealed sate, respectively. In the case of the as-deposited state, depicted from Fig. S9a to c, we confirmed that the EDS peaks for the V, V-$K_\alpha$ (4.952 keV), and V-$K_\beta$ (5.430 keV), were observed only in the Fig. S9b, which was captured from the beneath of CoFe area. The EDS spectra from the top and bottom of the structure, described in Fig. S9 and c, the trace of the V did not appear. In the case of 300°C annealed states, displayed in Fig. S9d-f, we could not attain the EDS peaks for the V at the top and bottom side of the structure where the normalized elemental line profile showed sizable intensity. Remarkably, we could capture the characteristic peaks for the V around the CoFe area in Fig. S9e. Therefore we concluded that the EDS line profiles for V, which were observed in the W/W-V and Ta area before and after 300°C annealing, were artifacts of the EDS measurement. Thus, the interdiffusion of the V atom was negligibly small.

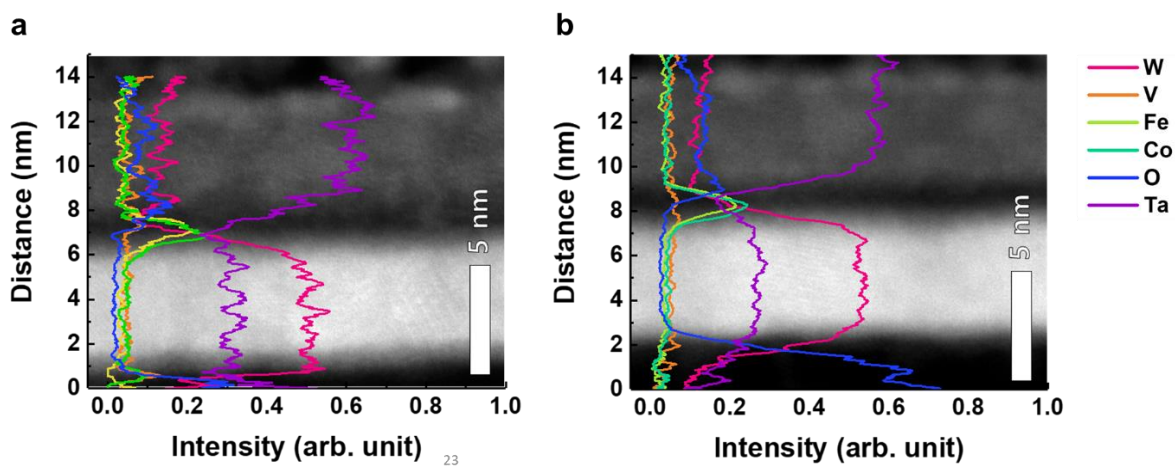

**Fig. S8 STEM data for the W/W-V/CoFeB/MgO/Ta structure. a,** HAADF-STEM image and elemental line profiles for the as-deposited state. **b,** HAADF-STEM image and elemental line profiles for 300°C annealed state.

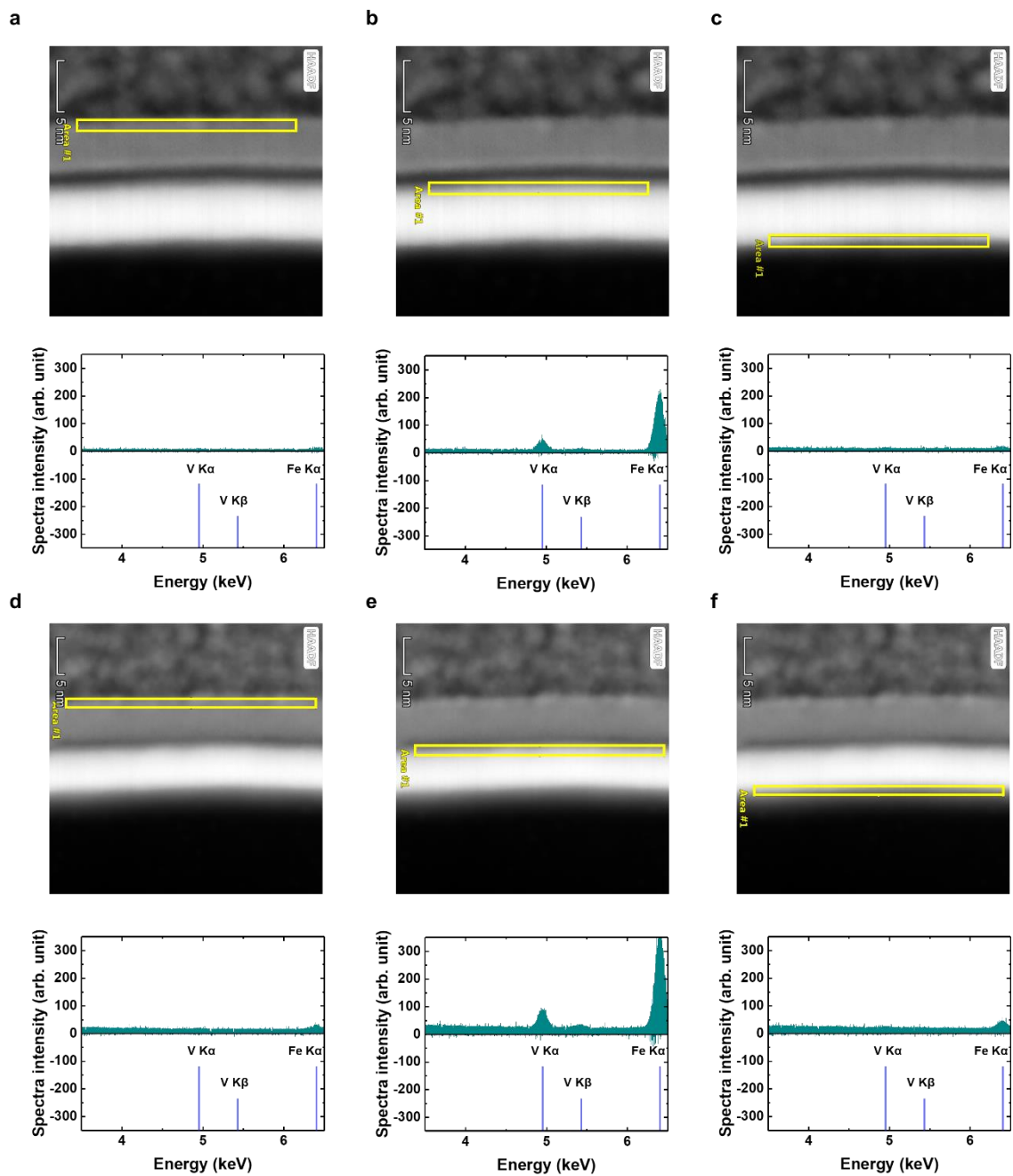

**Fig. S9 EDS spectra for a selected area. a-c,** HAADF-STEM image, which contains the location of the selected area and equivalent EDS spectra of as-deposited state. **d-f,** HAADF-STEM image, which includes the location of the selected area and identical EDS spectra of 300°C annealed state. The position of peaks for V-$K_\alpha$ (4.952 keV), V-$K_\beta$ (5.430 keV), and Fe-$K_\alpha$ (6.403 keV) are denoted in the EDS spectra.

**Note 9. Atomic distribution profiles in the W/W-X/CoFeB/MgO/Ta (X = V, Ta) structures**

To confirm whether the atomic interdiffusion occurred in the alloy component (V and Ta) in the film, we firstly measured the atomic profiles of the as-deposited and 300°C annealed samples by secondary ion mass spectroscopy (SIMS). According to Fig. S8, atomic distributions for both samples with W-V and W-Ta layers after 300°C annealing is almost identical to the as-deposited state.

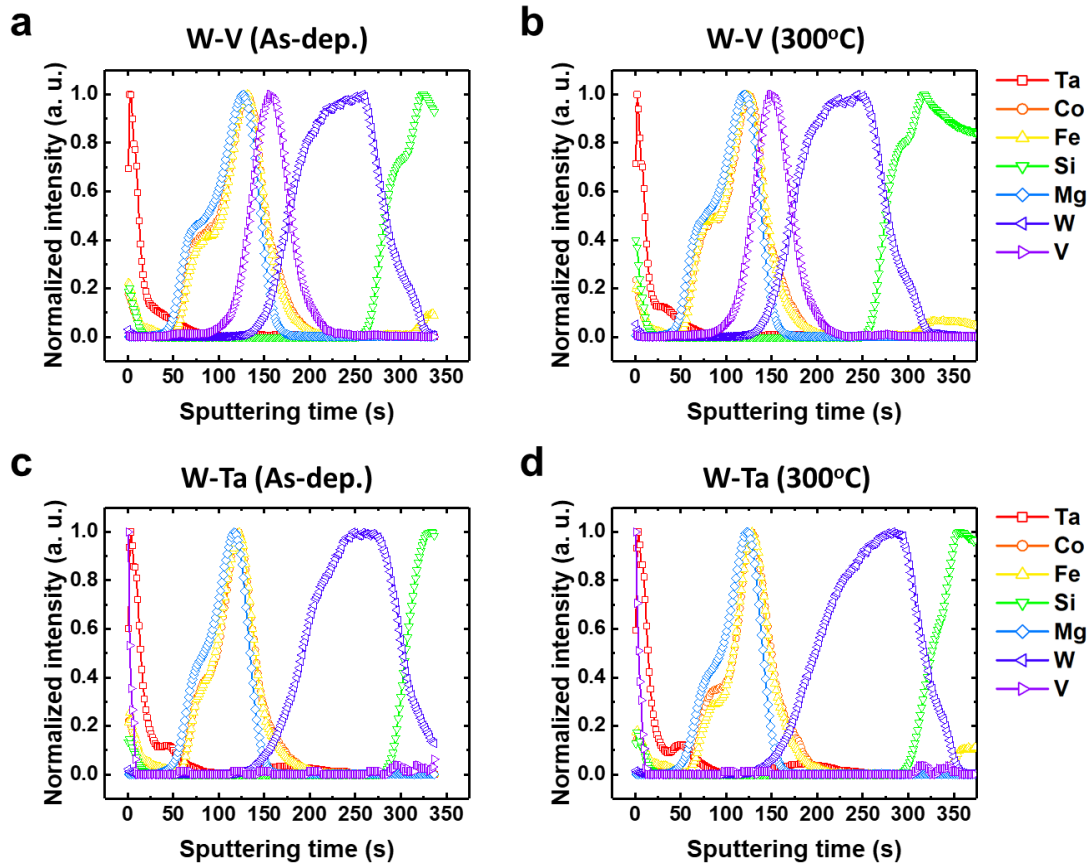

**Fig. S10 Secondary ion mass spectroscopy (SIMS) data for W/W-X/CoFeB/MgO/Ta (X = V, Ta) structure. a, b,** Normalized SIMS intensities of as-deposited 300°C annealed samples including interfacial W-V alloy layers. **c, d,** SIMS intensities of as-deposited 300°C annealed samples, including interfacial W-Ta alloy layers. The composition of the alloy layer is 50:50 in the atomic ratio.